\documentclass[%
 reprint,
 prl
 amsmath,amssymb,
 aps,
]{revtex4-1}

\usepackage{graphicx}
\usepackage{dcolumn}
\usepackage{bm}
\usepackage{amsmath}
\usepackage{xcolor}
\usepackage{url}

\newcommand{\mr}[1]{\mathrm{#1}}
\newcommand{\gt}{\gamma_\mr{t}}
\newcommand{\ds}{\Delta\Sigma}
\newcommand{\sigbar}{\overline{\Sigma}}
\newcommand{\sig}{\Sigma}
\newcommand{\rmax}{R_\mr{max}}

\newcommand{\ximm}{\xi_\mathrm{mm}}
\newcommand{\xigm}{\xi_\mathrm{gm}}

\newcommand{\hMpc}{h^{-1} {\rm Mpc}}

\begin{document}

\title{Localizing Transformations of the Galaxy--Galaxy Lensing Observable}

\author{Youngsoo Park}
\affiliation{Kavli Institute for the Physics and Mathematics of the Universe (WPI), UTIAS, The University of Tokyo, Kashiwa, Chiba 277-8583, Japan.}

\author{Eduardo Rozo}
\affiliation{Department of Physics, University of Arizona, AZ 85721, U.S.A.}

\author{Elisabeth Krause}
\affiliation{Department of Astronomy and Steward Observatory, University of Arizona, AZ 85721, U.S.A.\\
Department of Physics, University of Arizona, AZ 85721, U.S.A.}

\date{\today}

\begin{abstract}

Modern cosmological analyses of galaxy--galaxy lensing face a theoretical systematic effect arising from the non-locality of the observed galaxy--galaxy lensing signal. Because the predicted tangential shear signal at a given separation depends on the physical modeling on all scales internal to that separation, systematic uncertainties in the modeling of non-linear small scales are propagated outwards to larger scales. Even in the absence of other limiting factors, this systematic effect alone can necessitate conservative small-scale cuts, resulting in significant losses of information in the tangential shear data vector. We construct a simple linear transformation of the observable that removes this non-locality, enabling more aggressive small-scale cuts for a given theoretical model. Our modified galaxy--galaxy lensing observable makes it possible to include observations on significantly smaller scales than those under the standard approach in cosmological analyses. More importantly, it ensures that the cosmological signal contained within the observable is exclusively drawn from well-understood physical scales.
\end{abstract}

\maketitle

\section{Introduction}

Modern wide-field galaxy imaging surveys \citep[e.g.][]{desy1,2018MNRAS.474.4894J,kidsgama,hscwl,2013MNRAS.432.1544M} have achieved remarkable success in constraining cosmological parameters from measurements of the late-time matter distribution of our Universe. A promising avenue of recent interest reduces the measured positions and shapes of galaxies to two-point statistics in configuration space and compares those statistics against theoretical predictions for inference. There are three different types of two-point statistics that can be included in such analyses, namely position--position, position--shape, and shape--shape, respectively referred to as galaxy clustering, galaxy--galaxy lensing, and cosmic shear. When making use of these statistics, analysts must determine the range of scales that are modeled accurately enough for cosmological inferences, often in the form of scale cuts. On the small-scale side, the assumed model for the galaxy bias, which relates the spatial distribution of galaxies to that of matter \citep[see e.g.][for a review]{Desjacques:2016bnm}, limits the minimum possible scale that can be used in the analysis. For instance, a perturbative bias model will fail below some minimum physical scale, so comparisons between predictions and observations can only be made outside of that scale, even in the absence of other observational or theoretical systematic effects. 

For the case of galaxy--galaxy lensing, the minimum usable scale for analysis can be much larger than the limit discussed above. This is due to the \textit{non-local} nature of the galaxy--galaxy lensing signal, where the predicted signal at a given separation depends on the modeling of \textit{all} scales within that separation, including the non-linear small scales. In galaxy--galaxy lensing, the matter associated with a foreground lens galaxy tangentially distorts the images of background source galaxies. The raw observable is the azimuthally averaged ellipticity of background galaxies tangential to the lens-source separation vector, known as the tangential shear or $\gt$. For a lens at redshift $z_\mathrm{l}$ and a source at redshift $z_\mathrm{s}$, the tangential shear signal at an observed angular separation $\theta$, or equivalently at a projected comoving spatial separation $R=\theta\chi_\mathrm{l}$ with $\chi_\mathrm{l} = \chi(z_\mathrm{l})$ being the comoving distance to $z_\mathrm{l}$, is predicted to be
\begin{eqnarray}
\gt(\theta|z_\mathrm{l},z_\mathrm{s}) = \frac{\ds(R=\theta\chi_\mathrm{l})}{\Sigma_\mathrm{crit}(z_\mathrm{l},z_\mathrm{s})},
\label{eq:gtdef}
\end{eqnarray}
where $\Sigma_\mathrm{crit}(z_\mathrm{l},z_\mathrm{s})$ is a geometric factor given by
\begin{equation}
\Sigma_\mathrm{crit}(z_\mathrm{l},z_\mathrm{s}) = \frac{c^2}{4\pi G} \frac{(1+z)\chi_\mathrm{s}}{\chi_\mathrm{l}(\chi_\mathrm{s}-\chi_\mathrm{l})}
\end{equation}
with $\chi_\mathrm{s} = \chi(z_\mathrm{s})$, and $\ds$ being the excess surface density defined as
\begin{equation}
\ds(R) = \sigbar(<R) - \sig(R).
\label{eq:dsdef}
\end{equation}
Here, $\sig$ is the projected surface density, i.e.
\begin{eqnarray}
\sig(R) & = & \int_{-\infty}^{\infty} d R_\mathrm{z}\ \rho \left(\sqrt{R^2 + R_\mathrm{z}^2}\right) \nonumber \\
& = & \int_{-\infty}^{\infty} d R_\mathrm{z} \left[ 1 + \xigm\left(\sqrt{R^2 + R_\mathrm{z}^2}\right) \right],
\label{eq:sigdef}
\end{eqnarray}
of matter around the lens galaxy, with $\xi_\mathrm{gm}$ being the 3D galaxy-matter correlation function, and $\sigbar(<R)$ is the mean surface density internal to $R$, i.e.
\begin{equation}
    \sigbar(<R) = \frac{1}{\pi R^2} \int_0^R dR'~ 2\pi R' ~ \sig(R').
\label{eq:sigbardef}
\end{equation}
Note that while $\Sigma$ and $\Delta\Sigma$ implicitly depend on $z_\mathrm{l}$, we choose to suppress this in our notation for simplicity. Equations \ref{eq:dsdef}-\ref{eq:sigbardef} demonstrate that the lensing signal at a radius $R$, i.e. $\Delta\Sigma(R)$, depends on the surface mass density at all radii interior to $R$, i.e. $\Sigma(R')$ for $0 \leq R' \leq R$. Consequently, our inability to adequately model the small-scale density field necessarily impacts the lensing signal at {\it all} scales.  This {\it non-locality} can then force the small-scale cuts applied in real data to be significantly larger than the scale at which theoretical uncertainties in the density field become problematic. This was, for instance, what \cite{desy1mpp} found from their studies of theoretical systematics for the DES Y1 cosmology analysis.


\section{A Local Galaxy--Galaxy Lensing Observable}

In order to mitigate the non-locality in galaxy--galaxy lensing, one must remove the contribution to the galaxy--galaxy lensing observable from $\xigm(r)$ on small scales. Several such methods have been discussed, e.g. via subtracting out the estimated contribution from the mass enclosed within some cutoff scale \citep{upsilon_baldauf, upsilon_mandelbaum} or analytically marginalizing over the said contribution \citep{maccrann19}. In this work, we propose a novel approach that is motivated by the local quantity underlying the non-local signal, i.e. the projected surface density $\sig(R)$. 

The basic idea for our work is that since the surface density contrast $\Delta\Sigma$ is a linear transformation of the local surface density field $\Sigma$, inverting this relation would enable us to define an estimator of the surface density field, which would in turn be local.  To do so, we begin with the relation between the two:
\begin{eqnarray}
\ds(R) = \frac{1}{\pi R^2} \int_0^R dR'~ 2\pi R' ~ \sig(R') - \sig(R).
\end{eqnarray}
Differentiating each side with respect to R yields
\begin{eqnarray}
2\pi R \ds(R) + \pi R^2 \ds'(R) = -\pi R^2 \Sigma'(R),
\end{eqnarray}
and integrating again from $R$ to $\rmax$ yields
\begin{eqnarray}
\sig(R) & = & \sig(\rmax)  \nonumber \\
 & & + \int_R^{\rmax} dR'\
\left[ \frac{2\ds(R')}{R'} + \ds'(R') \right].
\label{eq:rec}
\end{eqnarray}
This is a well-known result \citep{kaiser95,johnston05} which demonstrates that one can reconstruct $\sig(R)$ from $\ds(R)$ up to an unknown constant $\sig(\rmax)$, typically referred to as the mass sheet degeneracy. 

However, we need not concern ourselves with reconstructing the exact $\sig(R)$ profile. Our goal here is simply to construct a local galaxy--galaxy lensing observable. To that end, we define a new quantity $Y$ via
\begin{eqnarray}
Y(R) & \equiv & \sig(R) - \sig(\rmax) \nonumber \\
& = & \int_R^{R_\mathrm{max}} d\ln R' ~ \left[2\ds(R') + \frac{d \ds (R')}{d \ln R'}\right].
\label{eq:defy}
\end{eqnarray}
From the first line, it is clear that $Y(R)$ is free of contributions from $\xigm$ on small ($r<R$) scales. From the second line we can see that $Y(R)$ can readily be calculated from $\ds(R')$ for $R\leq R' \leq \rmax$. Thus, $Y(R)$ is our desired observable.

In a real-life cosmology analysis, observables and predictions come in the form of discrete vectors rather than smooth functions. We thus proceed to discretize Equation \ref{eq:defy} as
\begin{eqnarray}
\mathbf{Y} & = & 2\mathcal{S}\mathbf{\ds} +\mathcal{SD} \mathbf{\ds} \nonumber \\
& = & (2\mathcal{S}+\mathcal{SD})\mathbf{\ds} \nonumber \\
& = & \mathcal{T}\mathbf{\ds},
\label{eq:tdef}
\end{eqnarray}
Here, $\mathcal{D}$ is the matrix of finite difference coefficients for discretized differentiation with respect to $\ln R$. To construct $\mathcal{D}$, we use central finite difference schemes of widths from 1 to 4 depending on the location of the point of interest. Where the central scheme cannot be used, i.e. for the first and the last row of $\mathcal{D}$, we use forward/backward schemes with width 4, respectively. $\mathcal{S}$ represents the trapezoidal summation matrix for discretized integration in $\ln R$. In light of the previous discussions, it is clear that $\mathcal{T}$ has the property of removing small-scale contributions from $\mathbf{\ds}$.  Just as importantly, since $\mathbf{Y}$ is a linear transformation of the original data vector, is is straightforward to compute the covariance matrix of the observable $\mathbf{Y}$,
\begin{equation}
\mathcal{C}_{\mathbf{Y}} = \mathcal{T} \mathcal{C}_{\mathbf{\ds}} \mathcal{T}^\textsf{T},
\label{eq:c1}
\end{equation}
where $\mathcal{C}_{\mathbf{\ds}}$ is the covariance matrix for $\mathbf{\ds}$.

These formulae for constructing $\mathbf{Y}$ and $\mathcal{C}_{\mathbf{Y}}$ can be used to transform any cosmology analysis utilizing $\ds$ as its galaxy--galaxy lensing observable into an analysis that utilizes $\mathbf{Y}$ as its observable.  In particular, given a prediction vector $\mathbf{\ds}_\mathrm{pred}$ and observation vector $\mathbf{\ds}_\mathrm{obs}$ with covariance matrix $\mathcal{C}_{\mathbf{\ds}}$, we can use $\mathcal{T}$ to obtain $\mathbf{Y}_\mathrm{pred}$, $\mathbf{Y}_\mathrm{obs}$, and $\mathcal{C}_{\mathbf{Y}}$. These transformed quantities can then be used for subsequent likelihood analyses in lieu of the original quantities of interest. Moreover, since $\gt$ is directly proportional to $\ds$ and $\mathcal{T}$ is a linear transformation, a similar procedure can be applied directly to $\gt$, i.e.
\begin{eqnarray}
\mathbf{Y}_\gamma & = & \mathcal{T} \mathbf{\gamma_\mathrm{t}}, \\
\mathcal{C}_{\mathbf{Y}_\gamma} & = & \mathcal{T} \mathcal{C}_{\mathbf{\gt}} \mathcal{T}^\textsf{T}.
\label{eq:c2}
\end{eqnarray}

\begin{figure*}[ht]
\includegraphics[width=\textwidth]{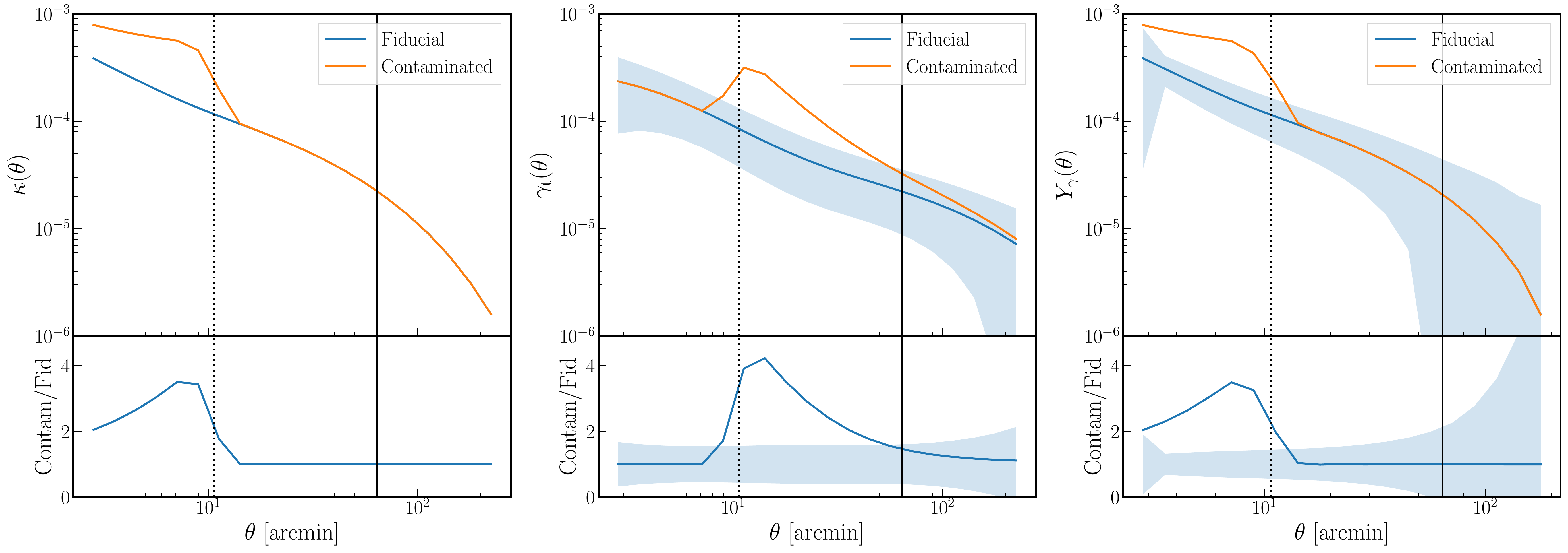}
\caption{Comparison of fiducial/contaminated data vectors for quantities $\kappa$ (left), $\gt$ (middle), and $\mathbf{Y}_\gamma$ (right). On each panel, we show the fiducial data vector constructed with linear bias and $\ximm$ from {\tt halofit} (blue) along with a contaminated data vector with an additional contribution from a disk-like mass profile on small scales (orange).  Note that the sharp radial boundary of the disk is smoothed over a range of angular scales due to the breadth of the lens redshift bins for the lenses.  The dotted and solid lines respectively represent $2$ and $12\ \hMpc$, i.e. $R_\mathrm{disk}$ and the DES Y1 $R_\mathrm{min}$, at the representative redshift of the fiducial lens sample. Starting from a localized underlying contamination in $\kappa$ (left), we observe the non-local propagation of the contamination in $\gt$ (middle), which is then successfully localized back by the linear transformation $\mathcal{T}$ of Equation \ref{eq:tdef} (right).}
\label{fig:local}
\end{figure*}

Now, it is clear that if one used the entire observable data vector $\gt$ or $\ds$, a linear transformation will not result in improved cosmological constraints.  However, the key point here is that all cosmological analyses impose a small scale cut, i.e. we only use a fraction of the observable data vector.  By applying the linear transformation derived above, we are able to apply a more aggressive small scale cut, and thereby preserve more of the information relative to a cut in $\gt$ or $\ds$.  In this context, it is also important to emphasize that the transformation matrix $\mathcal{T}$ must be ``extended'' for combined data vectors consisting of blocks other than galaxy--galaxy lensing, such that $\mathcal{T}$ would account for the full combined data vector. More specifically, we want $\mathcal{T}$ to transform the galaxy--galaxy lensing block while leaving the rest intact in such cases; this can be achieved by extending $\mathcal{T}$ block-diagonally with identity matrices that correspond to the data vector blocks outside of galaxy--galaxy lensing. Defining the transformation matrix across the full combined data vector is important since, in the presence of non-zero covariance between the galaxy--galaxy lensing block and other data blocks in the data vector, Equations~\ref{eq:c1} and \ref{eq:c2} will modify the off-diagonal components of the covariance matrix associated with the galaxy--galaxy lensing block.


\section{Testing the locality of $\mathbf{Y}$}

\begin{figure*}[ht]
\includegraphics[width=0.8\columnwidth]{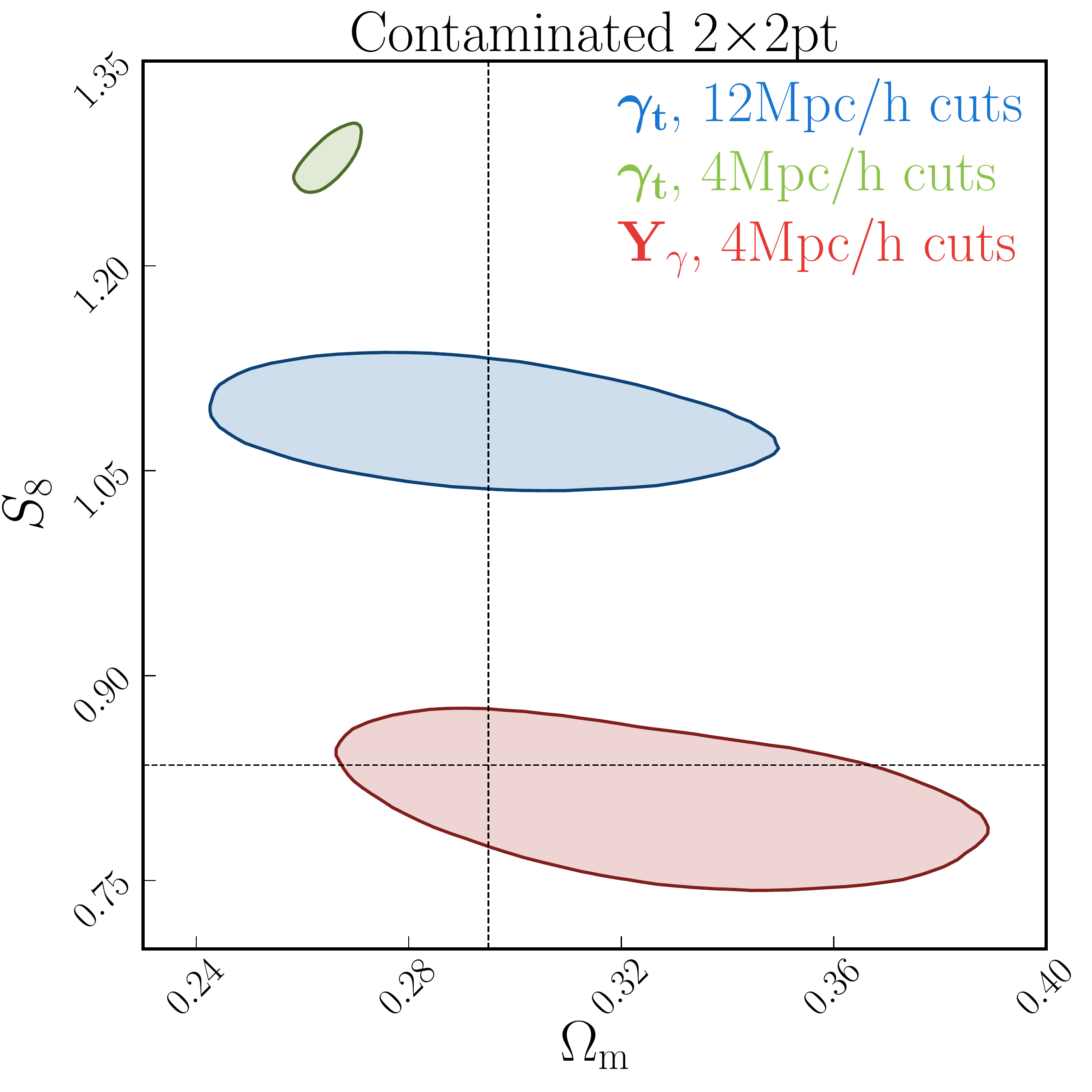}
\includegraphics[width=0.8\columnwidth]{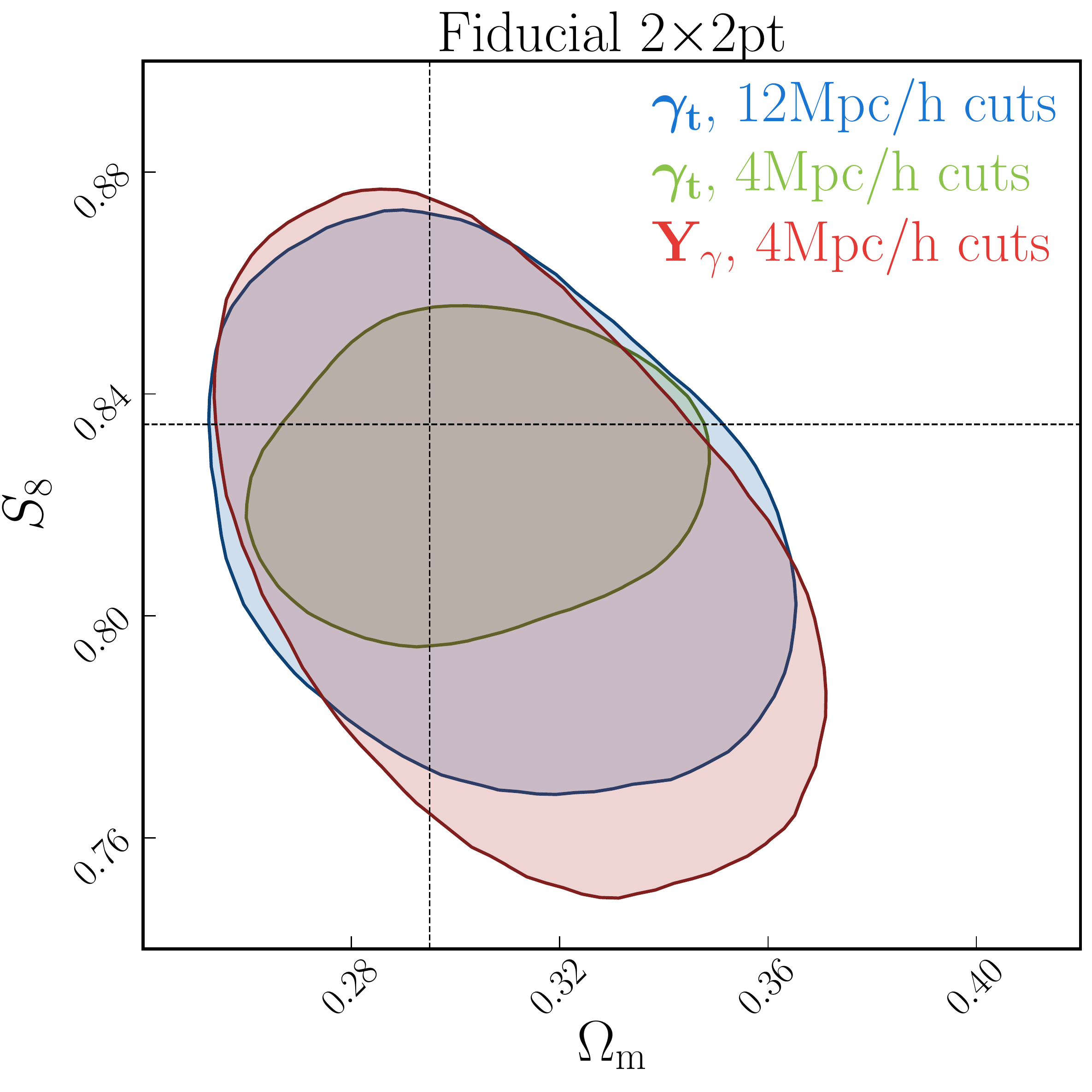}
\caption{68\% contours of the posterior in the $\Omega_{\rm m}$--$S_8$ plane for analyses employing the contaminated (left) and fiducial (right) data vectors. For each panel, the blue contours corresponds to the baseline analysis described in \cite{desy1}, using $\gt$ with a $12\ \hMpc$ small scale cut. The green and red contours correspond to analyses using $\gt$ with $4\ \hMpc$ cuts and using $\mathbf{Y}$ with $4\ \hMpc$ cuts, respectively. Dotted lines denote the fiducial parameter values used to generate the simulated data vectors. Note that the red contours, i.e. the posteriors derived from the $\mathbf{Y}$ observable proposed here, is nearly identical in both panels.}
\label{fig:contours}
\end{figure*}

We now test to what extent the new observable $Y$ is local, and whether the improved small scale cuts result in cosmological gains. The methodology for the simulated analyses, closely following \cite{desy1mpp}, is as follows. 
\begin{enumerate}
\item A simulated data vector is generated from theory with fiducial parameter values. The assumed theoretical model uses the {\tt halofit} \cite{halofit_takahashi} nonlinear matter power spectrum and linear galaxy bias.
\item The simulated data vector is ``contaminated'' with an unmodeled small-scale contribution from a disk-like mass distribution with a constant surface density.  The mass of the disk is $M$ and its radius is $R_\mathrm{disk}$. In terms of $\ds$, the contamination is given by
\begin{equation}
     \Delta\Sigma_\mathrm{disk}(R) =
     \begin{cases}
        0 & \text{for   } R \leq R_\mathrm{disk}\\
        M/\pi R^2 & \text{for   } R > R_\mathrm{disk}.
    \end{cases}
\end{equation}

We propagate this contamination to the $\gt$ observable, taking into account the lens and source redshift distributions used in \cite{desy1mpp}. We set $R_\mathrm{disk} = 2 \text{Mpc}/h$, and adopt masses 4 times larger than those used in \cite{desy1mpp}. While a disk contamination is not realistic, its sharp boundary best enables us to test whether the observable $Y$ displays non-locality.
\item The simulated contaminated data vector is analyzed without any modeling for the contamination, i.e. assuming the original theoretical model, in order to determine how cosmological inferences are impacted by this contamination.
\end{enumerate}
For all of the considered setups described below, we assume a ``2$\times$2pt'' cosmological analysis where the observables of interest are $\gt(\theta)$ (galaxy--galaxy lensing) and $w(\theta)$ (galaxy clustering). This combination consists of the minimal set of observables including galaxy--galaxy lensing that breaks the parameter degeneracy between galaxy bias and power spectrum amplitude. Note in particular that this analysis does not incorporate information from cosmic shear, as including cosmic shear will reduce the impact of contaminants in the galaxy--galaxy lensing data vector due to the increased ``clean'' cosmological information drawn from the cosmic shear signal.  

We make use of the public DES Y1 analysis pipeline \footnote{\protect\url{https://bitbucket.org/joezuntz/cosmosis/wiki/Home}} to generate simulated data vectors as well as theory predictions for these observables, and use the public DES Y1 covariances for likelihood calculation. The pipeline is also used to generate Markov Chain Monte Carlo (MCMC) samples for the likelihood analyses, from which we derive parameter constraints over the full DES Y1 parameter space. For more details, we refer the reader to \cite{desy1mpp}.  Here, we will be interested in the bias of the resulting cosmological constraints as the small scale cut used in the analysis of the galaxy--galaxy lensing signal becomes more aggressive.

In Figure \ref{fig:local}, we first show the localizing performance of the observable $\mathbf{Y}_\gamma$ by plotting the fiducial and contaminated observables for a single lens-source bin. The dotted and solid vertical lines respectively represent $R_\mathrm{disk}$ and the DES Y1 $R_\mathrm{min}$, i.e. 2 and 12 Mpc$/h$, in angular scales corresponding to the lens redshift distribution used. The left panel shows the local quantity $\kappa$, i.e. convergence, underlying the actual observable $\gt$. Note that the disk contamination is smoothed beyond $R_\mathrm{disk}$ due to the width of the lens redshift distribution.  The middle panel gives a clear illustration of the non-local behavior of $\gt$, where a mass distribution strictly limited to within $R_\mathrm{disk}$ contaminates the data vector well past the radius $R_\mathrm{disk}$.  The right panel demonstrates the localizing power of $\mathbf{Y}_\gamma$, where we observe that $\mathbf{Y}_\gamma$ can 1) correctly ``move back" the contamination to small scales and 2) recover the fiducial input to great accuracy on large scales.

Figure \ref{fig:contours} shows the 2D constraints in the $\Omega_\mathrm{m}$ -- $S_8$ plane from the simulated analyses, where $S_8 = \sigma_8 (\Omega_\mathrm{m}/0.3)^{0.5}$. The left panel clearly shows both the dangers of the non-locality in galaxy--galaxy lensing, as well as the power of using the localized observable $\mathbf{Y}_\gamma$. Due to the very large contamination employed in this test, even the baseline analysis of the galaxy--galaxy lensing signal $\gt$ with a conservative small scale cut of $12\ \hMpc$ shows highly biased results. The incurred bias is, unsurprisingly, much stronger for the $\gt$ analysis with $4\ \hMpc$ cuts. By contrast, we see that the use of the observable $\mathbf{Y}_\gamma$ results in unbiased cosmological posteriors down to the same small scale cut of $4\ \hMpc$. We stress that our method is agnostic to the type of contamination introduced. It is solely the localizing nature of $\mathbf{Y}_\gamma$ that is driving the debiasing exhibited in Fig. \ref{fig:contours}. 

The right panel shows the bias-variance tradeoff incurred by our approach. In particular, we now analyze an uncontaminated data vector for $\gt$ with $4\ \hMpc$ and $12\ \hMpc$ cuts, and compare it to a $\mathbf{Y}_\gamma$ data vector with a $4\ \hMpc$ scale cut. Note that the cosmological posteriors for the $\gt$ data vector with a $12\ \hMpc$ cut are tighter than the posteriors for the $\mathbf{Y}_\gamma$ data vector with a $4\ \hMpc$ cut.  Evidently, for a given scale cut, the $\mathbf{Y}_\gamma$ data vector contains significantly less information than the $\gt$ data vector.  Importantly, however, by construction, the $\mathbf{Y}_\gamma$ data vector should contain all of the information we can adequately model, and no information from scales that we are unable to properly model, a property that the $\gt$ data vector does not share.

\begin{figure}[t]
\includegraphics[width=\columnwidth]{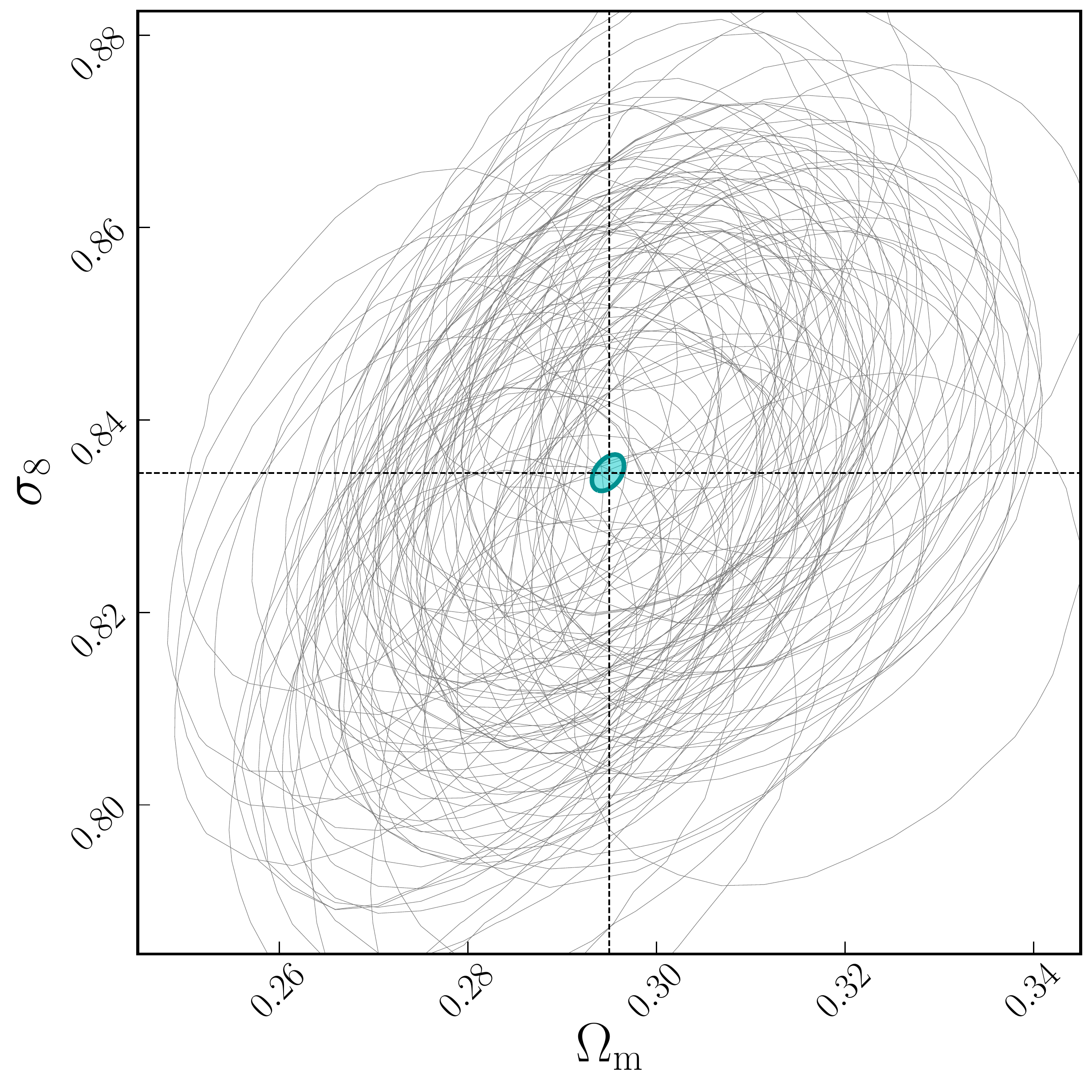}
\caption{Constraints on the $\Omega_\mathrm{m}$--$\sigma_8$ plane from simulated ``2$\times$2pt'' analyses utilizing $\mathbf{Y}_\gamma$, performed on 100 independent noisy realizations of the contaminated data vector. Each gray ellipse represents the resulting posterior for the 2D parameter space from a given noise realization, where both the galaxy--galaxy lensing and the galaxy clustering signals are perturbed according to the fiducial covariance matrix. The filled cyan ellipse represents the combined constraints from all 100 noise realizations, and is consistent with the fiducial parameter values marked by dotted lines.}
\label{fig:noise}
\end{figure}

Finally, we test our method for robustness against noise.  To do so, we analyze multiple independent noise realizations of the $\mathbf{Y}_\gamma$ data vector. We first generate 100 noisy realizations of the full fiducial data vector (i.e. $\gt$ and $w$) from the full fiducial covariance matrix, add the disk contamination to the $\gt$ section of each realization, and transform $\gt$ to $\mathbf{Y}_\gamma$. We then perform a grid-based likelihood analysis for each realization, fixing all cosmological and systematics parameters except for $\Omega_\mathrm{m}$ and $\sigma_8$. The analysis relies on the observable $\mathbf{Y}_\gamma$ with a $4\ \hMpc$ small scale cut.The results from the noise realization tests are shown in Fig. \ref{fig:noise}.  We find that the parameter constraints from independent noise realizations scatter around the fiducial parameter values, and in addition that the combined constraints from all 100 noise realizations are also consistent with the fiducial values. This shows that the transformations of the data vectors and covariances we introduce in our method are robust against random noise.

\section{Summary and Discussion}

We have introduced a novel observable to mitigate the non-locality in the galaxy--shear correlation function.  Our approach is inspired by the local quantity $\sig$ underlying the direct observables $\gt$ and $\ds$, and takes the form of a linear transformation on the observable vectors. By utilizing our localized observable $\mathbf{Y}$, we have obtained unbiased cosmological posteriors even under aggressive small-scale cuts in galaxy--galaxy lensing. Our approach is trivial to implement: starting from an existing analysis pipeline, one only needs to add a few matrix multiplications to the observables and the covariance matrices prior to computing the likelihood. Most importantly, our approach ensures that the entirety of the signal used to place cosmological constraints is free of non-local contributions from the small-scale regime, and consequently that the resulting cosmological constraints depend exclusively on accurately modeled physics.

\acknowledgements YP was partially supported by the DOE grant DE-SC0015975 and the World Premier International Research Center Initiative (WPI), MEXT, Japan, during the performance of this work. ER was supported by DOE grant DE-SC0015975, and by the Cottrell Scholar program of the Research Corporation for Science Advancement. EK was supported by DOE grant DE-SC0020247.

\newcommand\AAA{{A\& A}}
\newcommand\PhysRep{{Physics Reports}}
\newcommand\PhysRevD[3]{ {Phys. Rev. D}} 
\newcommand\jcap[3]{{JCAP}} 
\newcommand\PhysRevLet[3]{ {Phys. Rev. Letters} }
\newcommand\mnras{{MNRAS}}
\newcommand\PhysLet{{Physics Letters}}
\newcommand\AJ{{AJ}}
\newcommand\aap{ {A \& A}}
\newcommand\apjl{{ApJ Letters}}
\newcommand\aph{astro-ph/}
\newcommand\AREVAA{{Ann. Rev. A.\& A.}}
\newcommand\pasj{Publ. Aston. Soc. Jpn.}

\bibliography{papers}

\end{document}